\documentclass[%
 aip,
 amsmath,amssymb,
 reprint,%
]{revtex4-1}

\usepackage{graphicx}
\usepackage{dcolumn}
\usepackage{bm}

\usepackage[utf8]{inputenc}
\usepackage[T1]{fontenc}
\usepackage{mathptmx}

\begin{document}

\preprint{AIP/123-QED}

\title{Self-organized vortex and antivortex patterns in laser arrays}

\author{M. Honari-Latifpour}%
\author{J. Ding}
\author{S. Takei}
\author{M.-A. Miri}
\email{mmiri@qc.cuny.edu}
\affiliation{%
Department of Physics, Queens College of the City University of New York, Queens, New York 11367, USA\\
Physics Program, The Graduate Center, City University of New York, New York, New York 10016, USA
}%

\date{\today}

\begin{abstract}

Recently it is shown that dissipatively coupled laser arrays simulate the classical XY model. We show that phase-locking of laser arrays can give rise to the spontaneous formation of vortex and antivortex phase patterns that are analogous to topological defects of the XY model. These patterns are stable although their formation is less likely in comparison to the ground state lasing mode. In addition, we show that small ratios of photon to gain lifetime destabilizes vortex and antivortex phase patterns. These findings are important for studying topological effects in optics as well as for designing laser array devices.

\end{abstract}

\maketitle

\section{Introduction}

\noindent
The two-dimensional XY model consists of a lattice of interacting fixed-length spins that are constrained to rotate in a plane \cite{chaikin1995principles}. The classical XY model in two spatial dimensions is governed by the Hamiltonian $\mathcal{H}(\phi_1,\phi_2,\cdots,\phi_n)$:
\begin{equation}
\label{eq1}
    \mathcal{H}=\sum_{\langle i ,j \rangle}^{}\kappa_{ij}\cos(\phi_i-\phi_j)
\end{equation}
where, $\phi_1,\cdots,\phi_n$ represent the orientation of the $n$ spins, and $\kappa_{ij}$ is the interaction for the pair $i$,$j$. This model supports nontrivial equilibrium spin configurations~\textemdash~a class of topological defects~\textemdash~known as vortices, which are characterized by the phase of the spins going through a multiple of $2\pi$ as one traces a loop enclosing the vortex, e.g., 
\begin{equation}
    \int_{}^{} d\phi = \pm 2\pi
\end{equation}
for a single vortex/antivortex. In the 2D XY model, vortices have found applications in various areas of condensed matter physics including superfluid helium-4 \cite{bishop1980study}, superconductivity in thin films \cite{beasley1979possibility, hebard1980evidence}, liquid crystals \cite{pargellis1994planar}, and the melting of 2D crystals \cite{halperin1978theory, nelson1979dislocation}.

\begin{figure}
\flushleft
\includegraphics[width=0.481\textwidth]{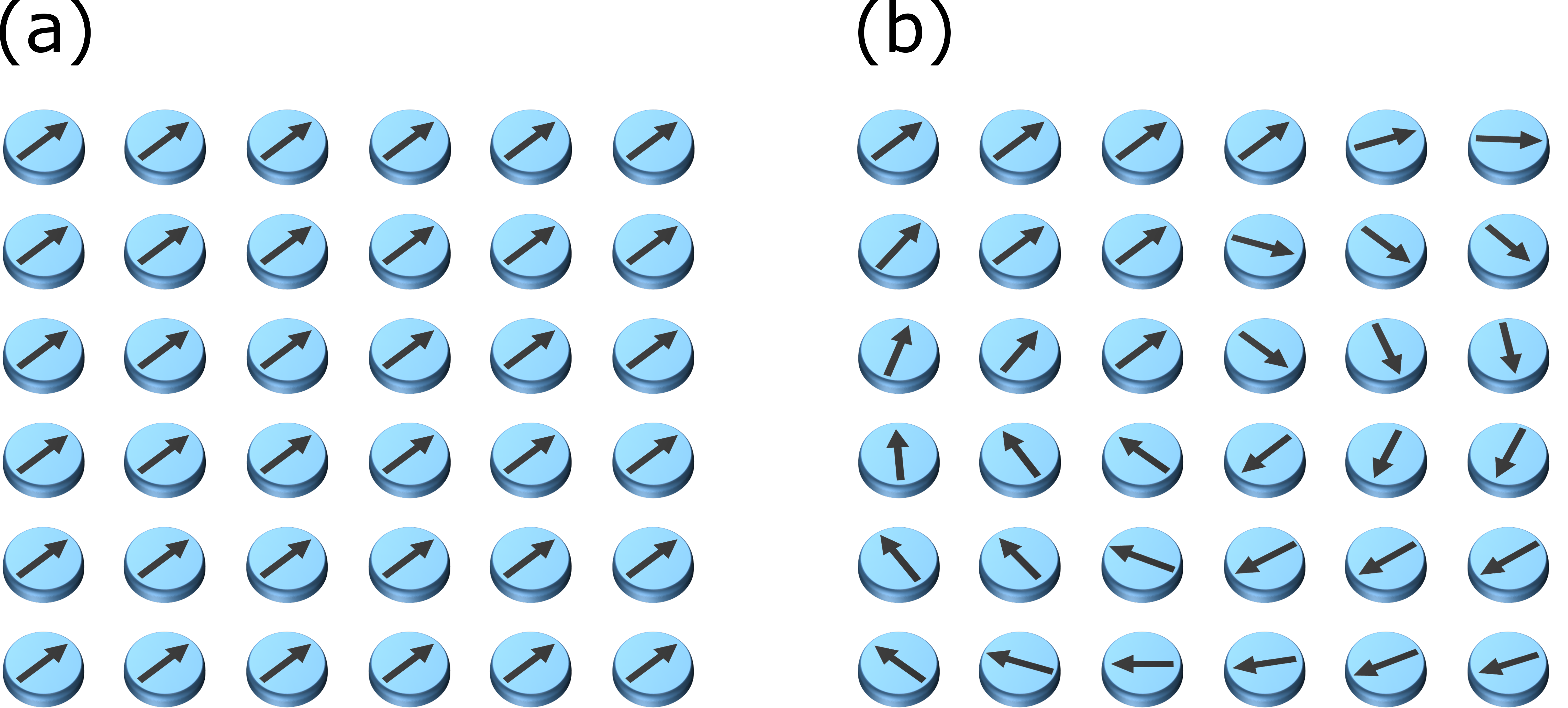}
\caption{A schematic of the equilibrium phase patterns of a dissipatively coupled laser array. (a) The ground state. (b) A vortex state. Here, the arrows represent the phase of each element.}
\label{fig1}
\end{figure}
Recently, it has been realized that the classical XY model can be optically simulated with an array of coupled optical oscillators \cite{nixon2013observing,berloff2017realizing,parto2020realizing,honari_2020}. What makes this possible is the random phase of a laser above oscillation thresholds, which emulates a classical spin confined to a two-dimensional plane. In addition, dissipative interaction facilitates synchronization of an array of lasers to a globally-phase-locked state \cite{ding2019dispersive}. In this case, one can show that the laser array reaches an equilibrium phase pattern that locally minimizes a cost function that is asymptotically equivalent with the classical XY Hamiltonian \cite{honari_2020}. Accordingly, laser arrays have been utilized for simulating interesting phenomena related to spin systems such as geometric frustration \cite{nixon2013observing}.

In this work, we investigate the formation of vortex and antivortex singularities as self-organized phase patterns in laser arrays. As shown schematically in Fig.~\ref{fig1}, these patterns are equilibrium states of laser arrays when reaching a globally-phase-locked state. From a nonlinear dynamics point of view, these topological defects are fixed-point solutions of nonlinear dynamical equations governing laser arrays. However, in general, they can be considered as metastable states with finite basins of attraction which restricts their formation to proper initial conditions. In addition, we find that the stability of these states depends critically on the gain level and lifetime. In the following, after introducing a dynamical model governing laser arrays, first we numerically investigate the formation of vortices. Next, we draw a connection between the governing dynamical model and a class of Ginzburg-Landau systems that are known to support vortex patterns. Finally, we investigate the stability of the vortex patterns with respect to the competing time scales of optical cavity and gain decay rates of the lasers.

\section{Model}

\begin{figure*}[ht]
\centering
\includegraphics[width=1\textwidth]{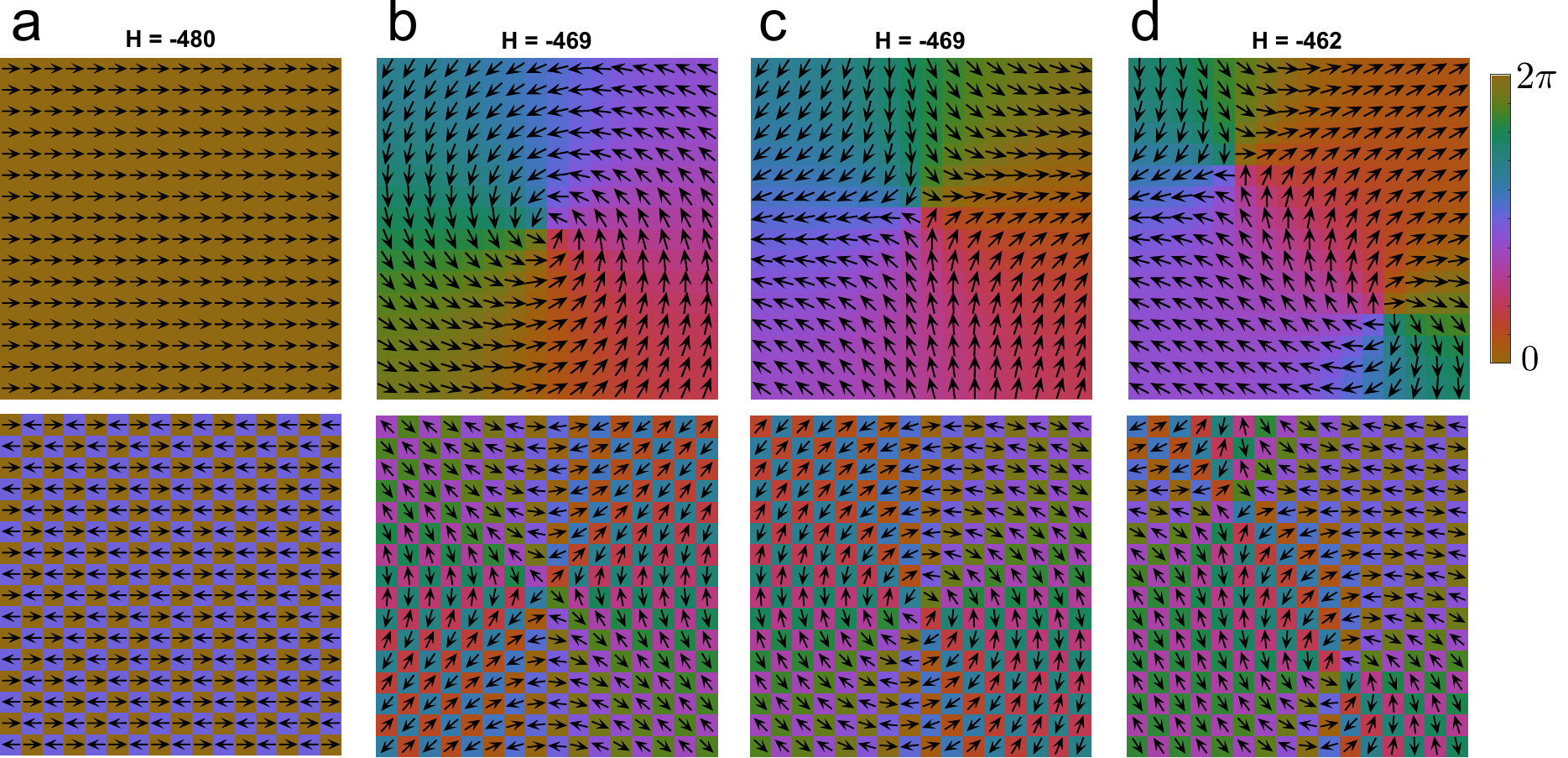}
\caption{Equilibrium phase patterns of an array of dissipatively coupled lasers arranged on a $16 \times 16$ square lattice. (a-d) The ground state, vortex, antivortex and bound vortex-antivortex states for the case of attractive coupling ($\kappa>0$), associated with a ferromagnetic spin system [first row], and repulsive coupling ($\kappa<0$), associated with an antiferromagnetic spin system [second row]. The XY energy level associated with these phase patterns are shown on top of the panels. Here, $g_0=30$, $\kappa=-1$ for the top row and $\kappa=1$ for the bottom row.}
\label{fig2}
\end{figure*}

To build a dynamical model governing laser arrays, we first consider an array of passive and single-mode optical resonators that dissipatively interact \cite{ding2019mode,ding2019dispersive}. For the sake of simplicity, we assume all resonators being identical in resonance frequency $\omega_0$ and linewidth $1/\tau_p$. Thus, in the framework of the temporal coupled mode theory \cite{haus1984waves}, the complex modal amplitude of the electric field in the $i$th resonator is governed by:
%
\begin{equation}
\label{eq3}
\dot{a}_{i} (t) =  - a_i - \gamma_i a_i - \sum _{j \neq i} \kappa_{ij} a_j ,
\end{equation}
%
where, the equations are written within the gauge $a_i \rightarrow a_i e^{-i\omega_0 t}$ and time is normalized to the photon lifetime $\tau_p$. In this relation, $\kappa_{ij}$ represents the rate of dissipative coupling between the $i$th and $j$th resonators, $\gamma_i = \sum_{j} |\kappa_{ij}|$ is the external loss of the $i$th resonator due to its coupling with other resonators as demanded by conservation relations \cite{suh2004temporal}. Here, all coupling coefficients $\kappa_{ij}$ are normalized to the photon decay rate $1/\tau_p$, while the choice of normalization for the complex amplitudes depends on the gain as discussed next.

\begin{figure}[b]
\flushleft
\includegraphics[width=0.48\textwidth]{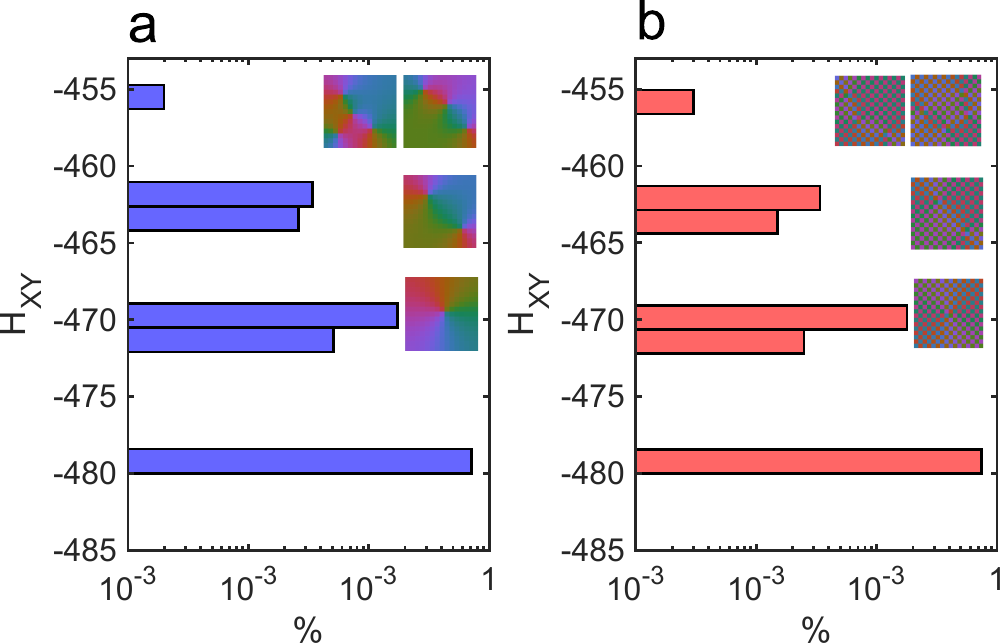}
\caption{The XY energy distribution associated with the equilibrium phase patterns of a $16 \times 16$ rectangular lattice laser array for (a) ferromagnetic, and (b) anti-ferromagnetic coupling. The histograms are produced based on $1000$ simulations with initial phases drawn randomly with uniform distribution from the range $[0,2\pi]$. All parameters are the same as in Fig.~\ref{fig2}.}
\label{fig3}
\end{figure}

By incorporating a saturable gain mechanism, equations~\eqref{eq3} can be modified to support self-sustained finite stationary solutions of the field amplitudes. Here, we consider the gain being a dynamical variable as in the so-called class-B laser model \cite{tredicce1985instabilities}. In this model, the gain of a laser oscillator is driven at a finite pump rate, while it decays linearly for small field intensities and nonlinearly when the field intensity grows. The normalized rate equations for the $i$th oscillator can then be written as: 
\begin{subequations}
\label{eq4}
\begin{align}
\dot{a}_{i} (t) & =  [g_i(t)-1] a_i - \gamma_i a_i - \sum _{j \neq i} \kappa_{ij} a_j , \\
\dot{g}_{i} (t) & =  (\tau_p/\tau_g) [g_0 - (1+|a_i|^2) g_i]  ,
\end{align}
\end{subequations}
where $g_i$ represents the gain of the $i$th oscillator, $g_0$ is the pump parameter, and $1/\tau_g$ is the gain decay rate. In these relations both the field amplitude and gain are dimensionless and the time is normalized to the photon lifetime $\tau_p$. This model has been applied to solid-state lasers \cite{fabiny1993coherence}. In addition, it can be generalized to model semiconductor lasers by incorporating the linewidth enhancement factor which plays an important role in the dynamics \cite{ohtsubo2012semiconductor}.

\begin{figure*}
\centering
\includegraphics[width=7in]{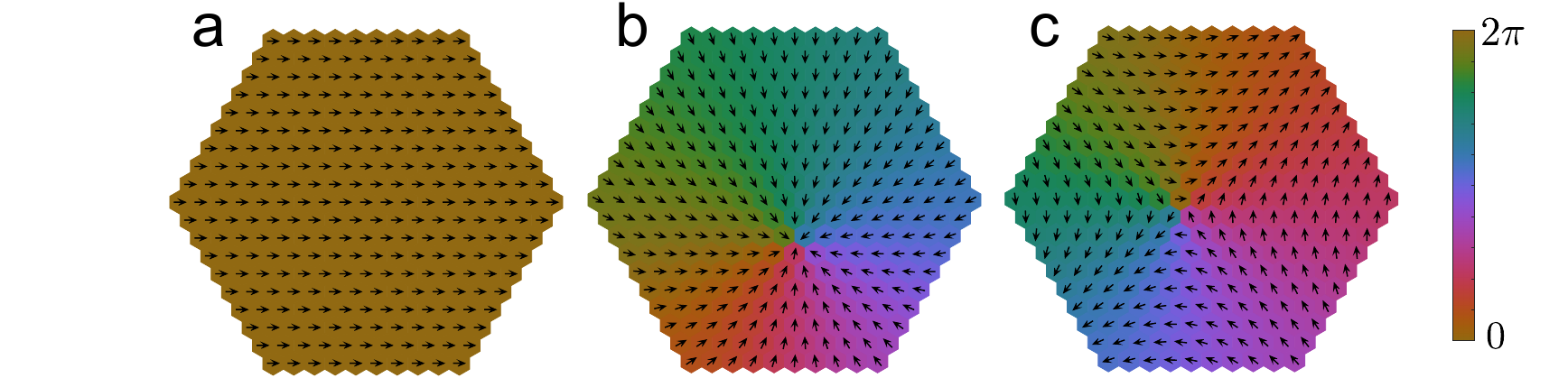}
\caption{The steady state phase pattern of a laser array arranged on a triangular lattice. Here, $\kappa=-1$ and $g_0=30$.}
\label{fig4}
\end{figure*}

\begin{figure}[b]
\flushleft
\includegraphics[width=0.48\textwidth]{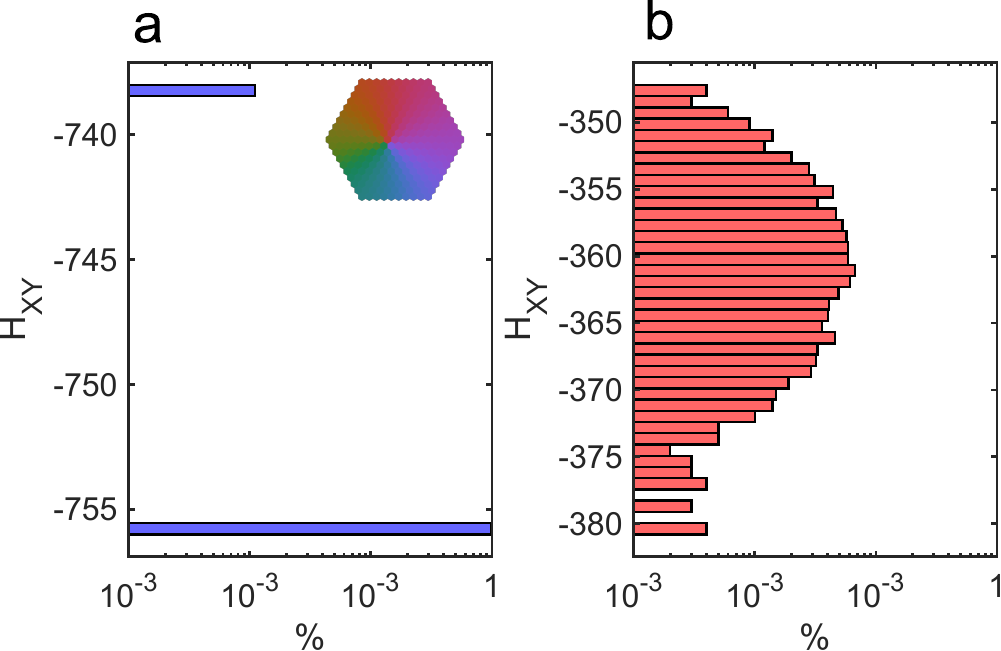}
\caption{The XY energy distribution associated with the equilibrium phase patterns of a triangular lattice arrangement of lasers with (a) ferromagnetic ($\kappa<0$), (b) anti-ferromagnetic ($\kappa>0$) coupling. The lattice size and array parameters are the same as in Fig.~\ref{fig4}}
\label{fig5}
\end{figure}

Equations~(\ref{eq4}) can be greatly simplified when the gain decay rate is much larger than the photon decay rate, i.e., $1/\tau_g \gg 1/\tau_p$. In this case, the gain almost instantaneously follows the dynamics of the field. Thus, one can adiabatically eliminate the dynamics of the gain, i.e., $\dot{g}_{i} (t) \approx 0$, to reach at an instantaneous nonlinear gain $g_i(|a_i|) = g_0/(1+|a_i|^2)$. In this manner, one reaches at a reduced model that is suitable for a so-called class-A laser \cite{tredicce1985instabilities,mandel1995optical}. Here, we use a polynomial gain term $g_i(a_i)=g_0(1-|a_i|^2)$, which is enough to guarantee the bounded oscillations of the laser. Thus, we reach at the following reduced model,
%
\begin{equation}
\label{eq5}
\dot{a}_{i} (t) =  [g_0 (1-|a_i|^2)-1] a_i - \gamma_i a_i - \sum _{j \neq i} \kappa_{ij} a_j, 
\end{equation}
%
which is accurate as long as the photon decay rate is smaller than the decay rates of the atomic degrees of freedom that give result to the gain. In this work, first we focus on simulating equations~(\ref{eq5}). Next, we return to equations~(\ref{eq4}) and discuss the effect of the gain dynamics.  

The dynamical model of Eq.~(\ref{eq5}) admits a Lyapunov function $F$ such that $\dot{a}_i = - \partial{F}/\partial{a_i^*}$, where \cite{honari_2020}
\begin{equation}
\label{eq6}
    F=\sum_{i}^{}[-(g_0-1-\gamma_i)|a_i|^2+\frac{g_0}{2}|a_i|^4]+ \frac{1}{2}\sum_{i,j}^{} a_i^*\kappa_{ij} a_j .
\end{equation}
The fixed point solutions of the dynamical system of Eq.~(\ref{eq5}) are local minima of this Lyapunov function. In a recent work we showed that the diagonal term in this cost function behaves like a penalty term that tends to force all oscillators to a constant amplitude in the large gain limit, $g_0 \gg 1$ \cite{honari_2020}. Thus, considering the stationary state solution of the oscillators as $a_i=\sqrt{I_i}e^{i \phi_i}$ by enforcing the condition of $I_i=I_0$, the cost function of Eq.~(\ref{eq6}) reduces to the XY Hamiltonian of Eq.~(\ref{eq1}).

\section{Formation of Topological Defects}

In the following, we investigate self-organization of topological defects by numerically simulating the dynamical model of Eqs.~(\ref{eq5}). We first consider an array of lasers arranged on a square lattice with uniform nearest-neighbor coupling of strength $\kappa$. Here, the gain is assumed to be large ($g_0 \gg 1$), so that  steady-state amplitudes become nearly uniform and the phase pattern obeys the XY Hamiltonian. Figure~\ref{fig2} depicts the equilibrium phase patterns obtained by simulating a square lattice arrangement of $16 \times 16$ oscillators for both scenarios of attractive ($\kappa<0$) and repulsive ($\kappa>0$) coupling, which are respectively associated with ferromagnetic and antiferromagnetic cases. In both cases, different stable patterns are observed, including the ground state, isolated vortex and antivortex states, and paired vortex-antivortex states. The XY energy levels associated with these equilibrium phase patterns are listed in Fig.~\ref{fig2}, which shows higher energy levels for the topological defects. The difference between the XY energy of the vortex (Fig.~\ref{fig2}(b)) and the ground state (Fig.~\ref{fig2}(a)) is comparable with the approximate formula $\Delta E = \pi \kappa \ln{L}$, where $L$ is the lattice length.

The stability of these fixed point solutions is directly evaluated through the Jacobian matrix of the dynamical system of Eqs.~(\ref{eq5}):
\begin{small}
\begin{equation}
\label{eqJ1}
\mathrm{J} =
%
%
g_0\begin{pmatrix}
\mathrm{diag}(1-2\bar{\mathbf{a}}\odot \bar{\mathbf{a}}^* ) - \mathrm{I} - \mathrm{Q} & -\mathrm{diag}(\bar{\mathbf{a}}\odot \bar{\mathbf{a}})  \\ -\mathrm{diag}(\bar{\mathbf{a}}^*\odot \bar{\mathbf{a}}^*) & \mathrm{diag}(1-2\bar{\mathbf{a}}\odot \bar{\mathbf{a}}^* ) - \mathrm{I} - \mathrm{Q} 
\end{pmatrix}.
\end{equation}
\end{small}

In this relation, $\bar{\mathbf{a}}=(\bar{a}_1,\cdots,\bar{a}_n)^t$ is the stationary state, $\mathrm{I}$ represents the identity matrix, $\mathrm{Q}$ is the coupling matrix, where $q_{ij}=\kappa_{ij}$ and $q_{ii}=\sum_{j} |\kappa_{ij}|$, $\odot$ represents the entry-wise product, and $\mathrm{diag}$ creates a diagonal matrix of a given vector. The Jacobian matrix turns out to be a negative semi-definite matrix for all cases shown in Fig.~\ref{fig2}.

\begin{figure*}
\flushleft
\includegraphics[width=1\textwidth]{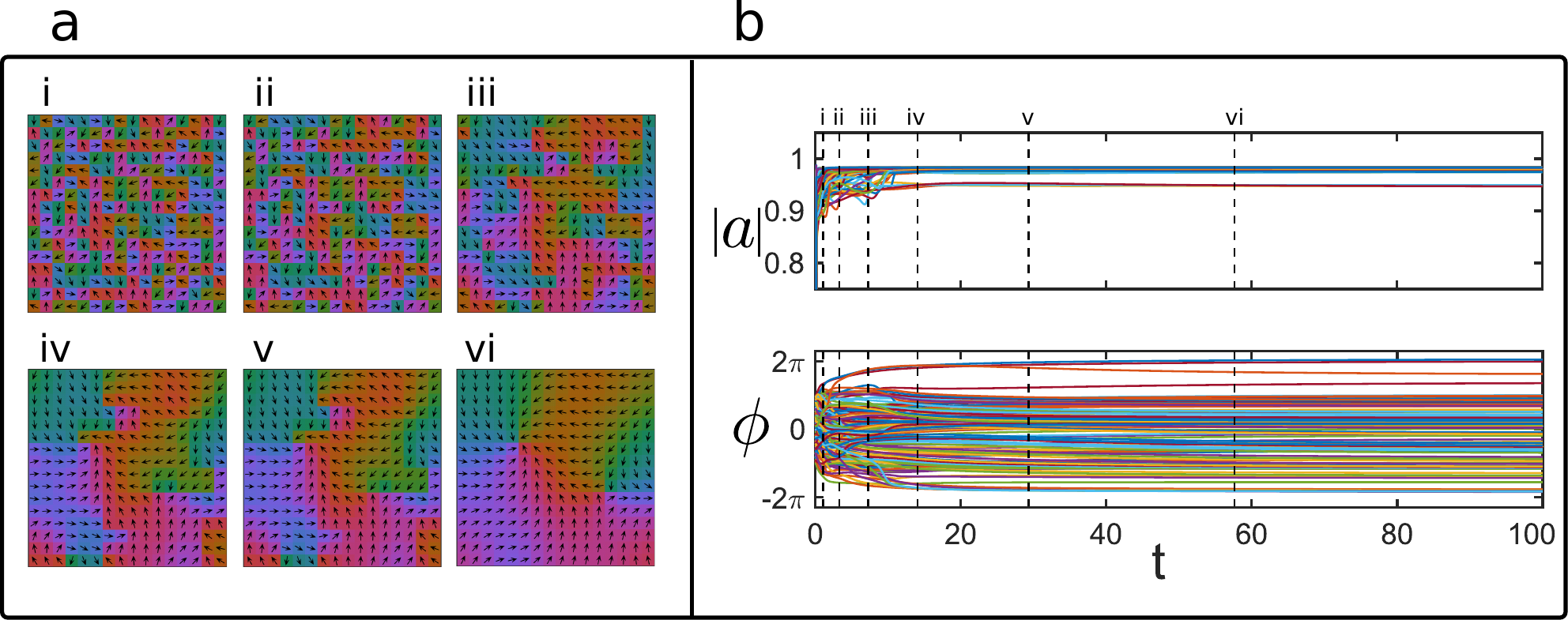}
\caption{The transient dynamics of a $16 \times 16$ laser array. (a) Snapshots of the phase pattern at intermediate times. (b) Amplitudes and and phases of the array elements. In part (b), the dashed lines show the time instants associated with the snapshots in part (a). All parameters are the same as in Fig.~\ref{fig2}.}
\label{fig6}
\end{figure*}

Given the finite attraction basins of the phase patterns shown in Fig.~\ref{fig2}, formation of these patterns depends on the initial conditions. In a realistic scenario, one can assume that the initial conditions are uniformly sampled from the high-dimensional phase space of the laser array. Thus, as local minima of the Lyapunov function, topological defects are expected to have lower chances of formation. To explore this aspect, we simulated the dynamical system for a large ensemble of random initial conditions. In these simulations, the initial phases were drawn from a uniform random distribution within the range $[0,2\pi]$, while the initial amplitudes were taken to be small values. The resulting equilibrium energy distribution is shown in Fig.~\ref{fig3} for both cases of ferromagnetic and antiferromagnetic systems. As this figure clearly indicates, topological defects form at much lower probabilities compared to the ground state.



The vortex and antivortex phase patterns can also form in the triangular lattice geometry as shown in the simulations of Fig.~\ref{fig4}. In this lattice geometry, the antiferromagnetic case is more complex due to the geometric frustration. The XY energy distributions associated with the equilibrium phase patterns of the ferromagnetic and antiferromagnetic cases are depicted in Fig.~\ref{fig5}. This figure indicates the complexity of the antiferromagnetic system due to its large number of higher energy states.

It is worth noting that the transient dynamics of the laser array reveals unstable topological defects that disappear by reaching the lattice boundaries or by collapsing of vortex-antivortex pairs. An exemplary dynamics of a laser network is shown in Fig.~\ref{fig6}. To visualize the transient dynamics, the phase pattern in depicted at intermediate time scales. It is worth noting that similar behavior was reported in a reduced model based on Kuramoto phase oscillators, which showed rapid decay of a large number of transient topological defects \cite{mahler2019dynamics}. 

\section{Discrete Ginzburg Landau Equation}

In order to provide insight to the formation vortices in laser arrays, we draw a connection between the lattice model of Eq.~(\ref{eq5}) with its continuum counterpart, which turns out to be the well-known Ginzburg-Landau Equation (GLE). To understand this analogy, we focus our attention to the case of a 2D square lattice arrangement of lasers with uniform nearest neighbor coupling $\kappa$, and consider the ferromagnetic case ($\kappa<0$). By considering a pair of integer indices $(m,n)$ for describing the horizontal and vertical coordinates of a square lattice, one can rewrite Eqs.~(\ref{eq5}) as:
\begin{equation}
\label{eq_dyn_square}
\dot{a}_{m,n} (t) =  [g_0(1-|a_{m,n}|^2)-1] a_{m,n} - \kappa \mathrm{L} a_{m,n}
\end{equation}
where, $\mathrm{L}$ is the discrete Laplacian operator on a square lattice graph that acts as $\mathrm{L} a_{m,n} = a_{m-1,n} + a_{m+1,n} + a_{m,n-1} + a_{m,n+1} - 4a_{m,n}$. This relation is clearly in a finite difference form. To construct the continuum counterpart of this relation, we use $(m,n) \rightarrow (x,y)$, $a_{m,n}(t) \rightarrow \psi(x,y,t)$, and $ \mathrm{L} \rightarrow \nabla^2$, which results in the Ginzburg-Landau Equation (GLE):
\begin{equation}
\label{eqGLE}
\dot{\psi}(x,y,t) =  [g_0(1-|\psi|^2)-1] \psi - \kappa \nabla^2 \psi .
\end{equation}
Given that all coefficients are real-valued, despite the fact that $\psi$ is complex, this equation is often referred to as the real Ginzburg-Landau equation \cite{aranson2002world}. 

\begin{figure*}
\flushleft
\includegraphics[width=1\textwidth]{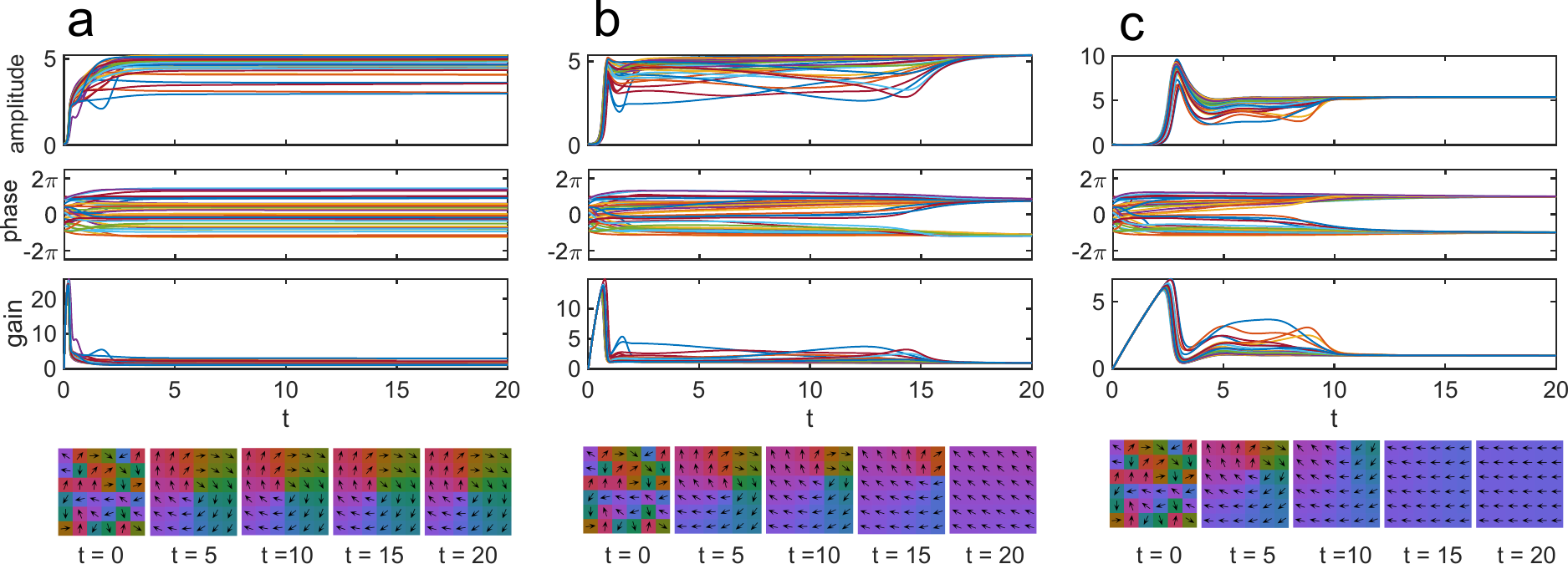}
\caption{The dynamics of a $6 \times 6$ array for (a) $\tau_p/\tau_g = 10$ (b) $1$, and (c) $0.1$. In all cases, $\kappa=-1$ and $g_0=30$.}
\label{fig7}
\end{figure*}

The dynamical equation~(\ref{eqGLE}) can be written in terms of a variation of a functional $\mathcal{F}[\psi]$, as $\dot{\psi} = - \delta \mathcal{F} / \delta \psi^*$. The functional $\mathcal{F}$ is found to be:
\begin{equation}
\label{eqGLEF}
\mathcal{F} = \int{dxdy \left[ - (g_0-1)|\psi|^2 + \frac{g_0}{2} |\psi|^4 + |\nabla \psi|^2 \right]} ,
\end{equation}
which, can be considered as a counterpart of the Lyapunov function of Eq.~(\ref{eq6}). In an amplitude and phase representation $\psi(r,\phi,t)=R(r,\phi,t) \exp[i\Phi(\phi(r,\phi,t))]$, Eq.~(\ref{eqGLE}) can be rewritten as:
\begin{subequations}
\label{eqLambdaOmega}
\begin{align}
\dot{R} (r,\phi,t) & = [g_0(1-R^2)-1] R - \kappa (\nabla^2 - |\nabla \Phi|^2 ) R, \\
R \dot{\Phi} (r,\phi,t) & = -\kappa ( 2\nabla R \cdot \nabla \Phi + R \nabla^2 \Phi).
\end{align}
\end{subequations}
These equations are a special class of a reaction-diffusion system, named $\lambda-\omega$ systems \cite{aranson2002world}. It is shown that these equations support stable single-arm spiral wave solutions of the form $R(r,\phi,t)=R(r)$ and $\Phi(r,\phi,t) = \phi $, while multi-arm spiral waves are unstable \cite{hagan1982spiral}. Driven from this intuition, one would expect discrete counterparts of the spiral patterns in lattices of coupled lasers. An obvious deviation of the lattice model from the continuum case is the finite amplitude of the field at the center of the vortex. In the continuum scenario, the vanishing amplitude at the phase singularity guarantees a well-defined field. In the case of the lattice model, however, there is no such requirement. This is clearly irrelevant in the case of the lattice model, although the amplitudes still tend to be lower at the lattice sites that are located near the vortex center. Nevertheless, the amplitudes become uniform in the large gain limit ($g_0 \rightarrow \infty$), where the laser system approaches the XY model.


\section{Vortices in the Dynamic Gain Regime}

The results presented in Figs.~\ref{fig2}-\ref{fig6} are based on the complex amplitude model with an instantaneous nonlinear saturable gain described by Eqs.~(\ref{eq5}). Here, we consider the more realistic scenario that involves gain as a dynamical variable as described through Eqs.~(\ref{eq4}). The gain lifetime $\tau_g$, which in reality is dictated by the laser gain material, is found to play a critical role in the dynamics and in formation of stationary topological defects. As mentioned earlier, when the ratio of the photon lifetime over the gain lifetime is large ($\tau_p/\tau_g \gg 1$), the gain dynamics can be effectively disregarded. Thus, in this regime, stable vortex and antivortex patterns are expected. It remains to investigate the stability of these topological defects when the photon to gain lifetime ratio becomes small ($\tau_p/\tau_g \sim 1$ or $\tau_p/\tau_g \ll 1$).

To explore this aspect, we first simulated the dynamics of a small array of lasers for different values of $\tau_p/\tau_g$ while using identical initial conditions. The results are depicted in Figs.~\ref{fig7}. As these exemplary simulations indicate, by decreasing the ratio of $\tau_p/\tau_g$, the transient dynamics becomes more complicated, involving effects like self-pulsations that are known processes in class-B lasers. On the other hand, these figures show that the vortex state does not form for smaller ratios of $\tau_p/\tau_g$ (Figs.~\ref{fig7}(b,c)). 

\begin{figure}
\centering
\includegraphics[width=3.3in]{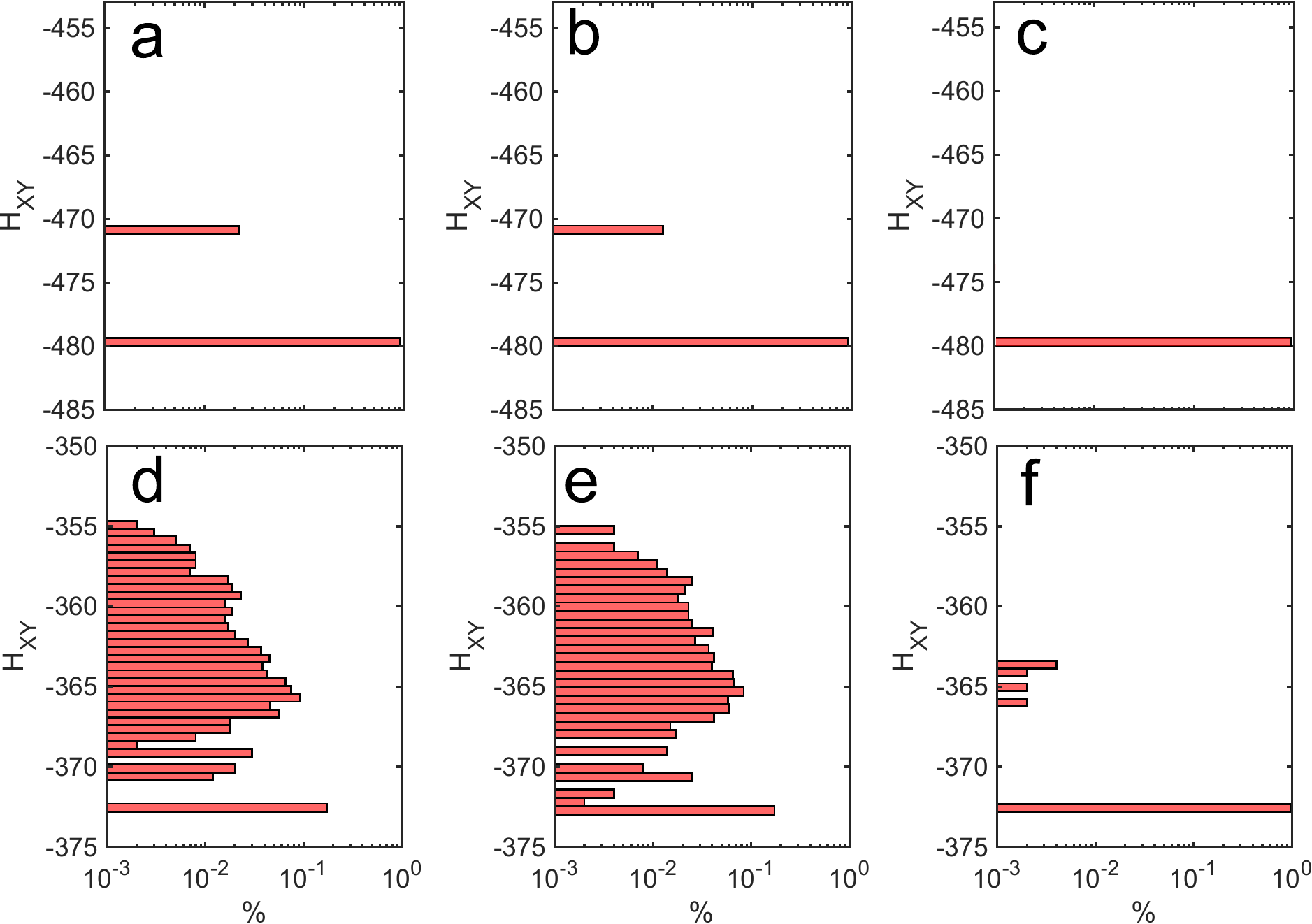}
\caption{The XY energy distribution associated with the equilibrium phase pattern of laser arrays governed by the dynamical equations (\ref{eq4}) for different photon to gain lifetime ratios. Top: The square lattice for (a) $\tau_p/\tau_g = 10$, (b) $1.0$, and (c) $0.01$. Bottom: The triangular lattice for (a) $\tau_p/\tau_g = 10$, (b) $1.0$, and (c) $0.01$. Here, $\kappa = -1$ and $g_0 = 30$, while the lattice size is the same as in Figs.~\ref{fig2} and \ref{fig4} for the top and bottom rows, respectively.}
\label{fig8}
\end{figure}

This can be explained through the dynamics of the gain. According to Figs.~\ref{fig7}(b,c), by increasing $\tau_g$, the gain increases slowly and during this process, the laser array finds the time to escape from patterns associated with higher XY energy levels and to stabilize into the ground state phase pattern. This is in agreement with our recent study based on the reduced model, while the gain was adiabatically increased to avoid trapping into local minima \cite{honari_2020}. Here, the dynamics allows the gain to automatically increase over a time scale governed by $\tau_g$. To systematically explore this aspect, the laser array was simulated with an ensemble of random initial conditions and for three different time scale ratios. In addition, both cases of rectangular and triangular lattices were considered. The XY energy distributions of the associated steady-state phase patterns are shown in Fig.~\ref{fig8}. As expected, for $\tau_p/\tau_g \gg 1$, the equilibrium energy distribution is similar to the case of instantaneous gain (Figs.~\ref{fig3}(b) and \ref{fig5}(b)). As this ratio increases, however, the occurrence of higher energy patterns drops significantly.

\begin{figure}
\centering
\includegraphics[width=3in]{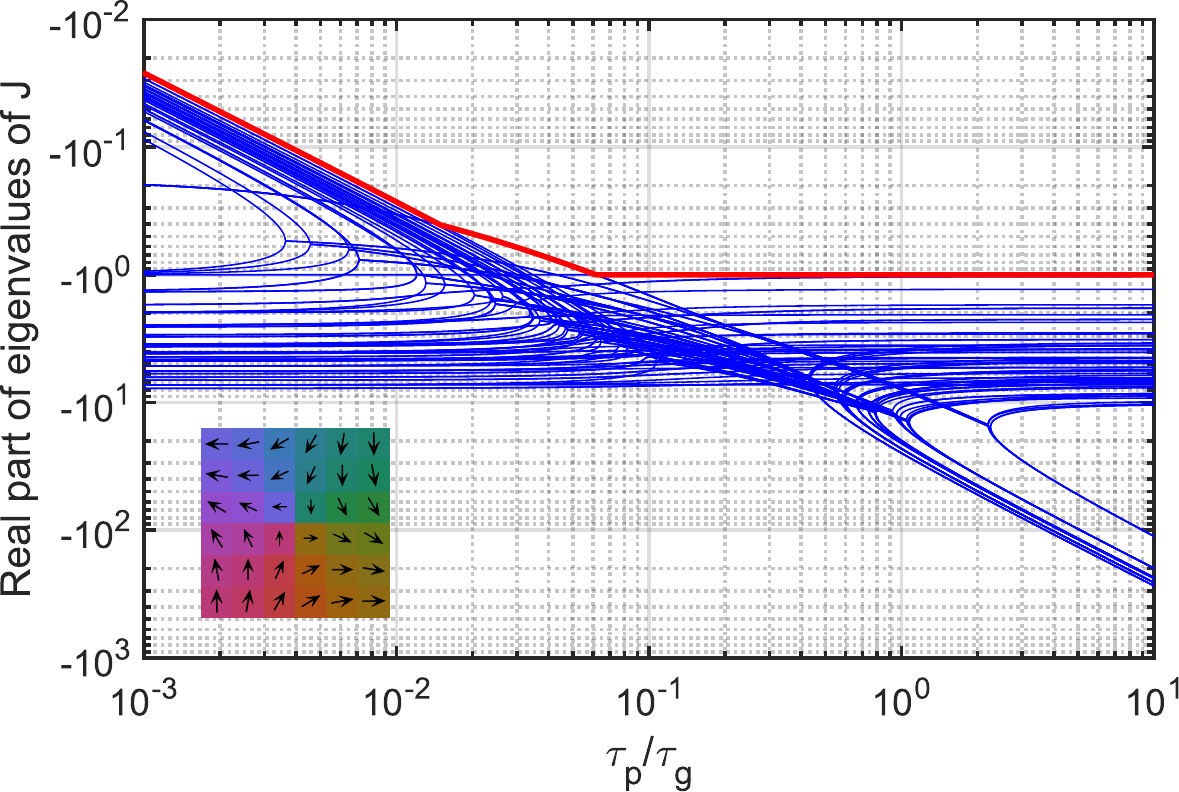}
\caption{The real part of the Jacobian matrix eigenvalues versus the photon to gain lifetime ratio $\tau_p/\tau_g$. These eigenvalues represent the linear stability test for a vortex pattern (shown as inset) in a $6 \times 6$ rectangular lattice array. The largest eigenvalue is highlighted in red.}
\label{fig9}
\end{figure}

These results clearly suggest that smaller ratios of $\tau_p/\tau_g$ destabilize the vortex and antivortex patterns. However, it is quite interesting to note that these patterns pass the linear stability test in the both regimes of small and large time scale ratios. To show this, first we consider the fixed point solutions of Eqs.~(\ref{eq4}), i.e., $(a_i(t),g_i(t))=(\bar{a}_i,\bar{g}_i)$. The fixed points are governed by a set of nonlinear algebraic equations:
\begin{subequations}
\label{eq:dyn1}
\begin{align}
[\bar{g}_i-1] \bar{a}_i - \gamma_i \bar{a}_i - \sum _{j \neq i} \kappa_{ij} \bar{a}_j = 0 , \\
[g_0 - (1+|\bar{a}_i|^2) \bar{g}_i] = 0 ,
\end{align}
\end{subequations}
which, are independent from the time scale ratio $\tau_p/\tau_g$. Thus, as an example, the vortex state found in Fig.~\ref{fig7}(a) for $\tau_p/\tau_g = 10$ is also a valid solution of the system explored in Fig.~\ref{fig7}(c) for $\tau_p/\tau_g = 0.1$. However, while numerical simulations indicate that the vortex state is stable in the former case, it remains to investigate its stability in the latter scenario. In this case again, it is straightforward to perform a linear stability analysis through the Jacobian matrix:
%
%
%
%
%
%
%
%
%
%
\begin{small}
\begin{equation}
\label{eqJ2}
\mathrm{J} =
\begin{pmatrix}
\mathrm{diag}(\bar{\mathbf{g}}) - \mathrm{I} - \mathrm{Q} & \mathrm{diag}(\bar{\mathbf{a}}) & \mathbf{0} \\ 
-\tau \mathrm{diag}(\bar{\mathbf{a}} \odot \bar{\mathbf{g}})^* & -\tau (\mathrm{I}+\mathrm{diag}(\bar{\mathbf{a}} \odot \bar{\mathbf{a}}^*)) & -\tau \mathrm{diag}(\bar{\mathbf{a}} \odot \bar{\mathbf{g}})\\ 
\mathbf{0} & \mathrm{diag}(\bar{\mathbf{a}})^* & \mathrm{diag}(\bar{\mathbf{g}}) - \mathrm{I} - \mathrm{Q}
\end{pmatrix}
\end{equation}
\end{small}
where, $\bar{\mathbf{a}}=(\bar{a}_1,\cdots,\bar{a}_n)^t$, $\bar{\mathbf{g}}=(\bar{g}_1,\cdots,\bar{g}_n)^t$, and $\tau=\tau_p/\tau_g$. By considering the vortex pattern of Fig.~\ref{fig7}(a) as the fixed point solution, the eigenvalues of the Jacobian matrix were numerically calculated for a range of values of $\tau_p/\tau_g$. Figure~\ref{fig9} depicts the real part of all eigenvalues versus $\tau_p/\tau_g$, while the lead eigenvalue, i.e., the one with the largest real part, is highlighted in red. Clearly, all eigenvalues exhibit negative real parts in the entire range of the time scale ratio, which shows linear stability of the vortex state in all regimes. However, according to this figure, the eigenvalues undergo a square-root transition and increase toward zero for small values of $\tau_p/\tau_g$. This clearly suggests the evolution of the vortex from a stable to a metastable state that can disappear with a small perturbation, which is in agreement with the simulations.  

\section{Conclusion}

In summary, we investigated the formation of vortex and antivortex phase patterns in laser arrays. We showed that for large gains these patterns exist in direct analogy with topological defects of the XY model. However, their stability depends critically on the ratio of photon to gain lifetime $\tau_p/\tau_g$. An important finding was that large gain lifetimes ($\tau_g$) destabilize topological defects and force the system into the ground state.

The presence of self-organized vortex and antivortex phase patterns in laser arrays could have practical applications given that these patterns can emit optical vortex beams. It is worth noting that both solid-state and semiconductor lasers are considered class-B lasers given that in these systems the fluorescence lifetime is by several orders of magnitudes larger than the photon lifetime ($\tau_p/\tau_g \ll 1$). A potential route for enforcing such laser arrays to create vortex phase patterns is to fix the phase of the boundary elements which could be done by seeding through a master laser. In this manner, vortex patterns can be enforced by fixing the topological charge of the array through boundary elements. However, it remains to investigate this aspect given that pinning the topological charge in the nonlinear dynamical systems discussed here can result in wandering of a vortex and potentially other instabilities. 

Finally, it is worth mentioning that the formation of bound vortex-antivortex states suggests an analogy with the Berezinskii–Kosterlitz–Thouless (BKT) phase transition. In two spatial dimensions (2D), the theorem by Mermin and Wagner precludes conventional long-range order at any finite temperature in systems with continuous symmetry and short-range interactions \cite{mermin1966absence}. Nevertheless, certain 2D systems can exhibit signs of a BKT transition from a high-temperature disordered state, with short-range correlations, to a quasi-ordered state, with algebraic correlations, below a critical temperature\cite{kosterlitz1973j, kosterlitz1974critical}. When viewed in terms of the vortices, the BKT transition is a vortex-unbinding transition. These topological objects, while tightly bound in pairs in the low-temperature state, unbind and freely proliferate above the critical temperature, causing the system to melt its quasi-order.  While the present work suggests such a transition, a rigorous investigation of this aspect requires a non-equilibrium thermodynamic treatment of the laser array which could be the subject of future studies.
\begin{acknowledgments}
We thank Dr. Morrel H. Cohen for bringing the subject of the BKT phase transition to our attention.
\end{acknowledgments}
\noindent
\section*{DATA AVAILABILITY}
The data that support the findings of this study are available from the corresponding author upon reasonable request.
\noindent
\section*{references}
\nocite{*}
\bibliography{aipsamp}

\begin{thebibliography}{26}%
\makeatletter
\providecommand \@ifxundefined [1]{%
 \@ifx{#1\undefined}
}%
\providecommand \@ifnum [1]{%
 \ifnum #1\expandafter \@firstoftwo
 \else \expandafter \@secondoftwo
 \fi
}%
\providecommand \@ifx [1]{%
 \ifx #1\expandafter \@firstoftwo
 \else \expandafter \@secondoftwo
 \fi
}%
\providecommand \natexlab [1]{#1}%
\providecommand \enquote  [1]{``#1''}%
\providecommand \bibnamefont  [1]{#1}%
\providecommand \bibfnamefont [1]{#1}%
\providecommand \citenamefont [1]{#1}%
\providecommand \href@noop [0]{\@secondoftwo}%
\providecommand \href [0]{\begingroup \@sanitize@url \@href}%
\providecommand \@href[1]{\@@startlink{#1}\@@href}%
\providecommand \@@href[1]{\endgroup#1\@@endlink}%
\providecommand \@sanitize@url [0]{\catcode `\\12\catcode `\$12\catcode
  `\&12\catcode `\#12\catcode `\^12\catcode `\_12\catcode `\%12\relax}%
\providecommand \@@startlink[1]{}%
\providecommand \@@endlink[0]{}%
\providecommand \url  [0]{\begingroup\@sanitize@url \@url }%
\providecommand \@url [1]{\endgroup\@href {#1}{\urlprefix }}%
\providecommand \urlprefix  [0]{URL }%
\providecommand \Eprint [0]{\href }%
\providecommand \doibase [0]{http://dx.doi.org/}%
\providecommand \selectlanguage [0]{\@gobble}%
\providecommand \bibinfo  [0]{\@secondoftwo}%
\providecommand \bibfield  [0]{\@secondoftwo}%
\providecommand \translation [1]{[#1]}%
\providecommand \BibitemOpen [0]{}%
\providecommand \bibitemStop [0]{}%
\providecommand \bibitemNoStop [0]{.\EOS\space}%
\providecommand \EOS [0]{\spacefactor3000\relax}%
\providecommand \BibitemShut  [1]{\csname bibitem#1\endcsname}%
\let\auto@bib@innerbib\@empty
\bibitem [{\citenamefont {Chaikin}, \citenamefont {Lubensky},\ and\
  \citenamefont {Witten}(1995)}]{chaikin1995principles}%
  \BibitemOpen
  \bibfield  {author} {\bibinfo {author} {\bibfnamefont {P.~M.}\ \bibnamefont
  {Chaikin}}, \bibinfo {author} {\bibfnamefont {T.~C.}\ \bibnamefont
  {Lubensky}}, \ and\ \bibinfo {author} {\bibfnamefont {T.~A.}\ \bibnamefont
  {Witten}},\ }\href@noop {} {\emph {\bibinfo {title} {Principles of condensed
  matter physics}}},\ Vol.~\bibinfo {volume} {10}\ (\bibinfo  {publisher}
  {Cambridge university press Cambridge},\ \bibinfo {year} {1995})\BibitemShut
  {NoStop}%
\bibitem [{\citenamefont {Bishop}\ and\ \citenamefont
  {Reppy}(1980)}]{bishop1980study}%
  \BibitemOpen
  \bibfield  {author} {\bibinfo {author} {\bibfnamefont {D.}~\bibnamefont
  {Bishop}}\ and\ \bibinfo {author} {\bibfnamefont {J.}~\bibnamefont {Reppy}},\
  }\bibfield  {title} {\enquote {\bibinfo {title} {Study of the superfluid
  transition in two-dimensional he 4 films},}\ }\href@noop {} {\bibfield
  {journal} {\bibinfo  {journal} {Physical Review B}\ }\textbf {\bibinfo
  {volume} {22}},\ \bibinfo {pages} {5171} (\bibinfo {year}
  {1980})}\BibitemShut {NoStop}%
\bibitem [{\citenamefont {Beasley}, \citenamefont {Mooij},\ and\ \citenamefont
  {Orlando}(1979)}]{beasley1979possibility}%
  \BibitemOpen
  \bibfield  {author} {\bibinfo {author} {\bibfnamefont {M.}~\bibnamefont
  {Beasley}}, \bibinfo {author} {\bibfnamefont {J.}~\bibnamefont {Mooij}}, \
  and\ \bibinfo {author} {\bibfnamefont {T.}~\bibnamefont {Orlando}},\
  }\bibfield  {title} {\enquote {\bibinfo {title} {Possibility of
  vortex-antivortex pair dissociation in two-dimensional superconductors},}\
  }\href@noop {} {\bibfield  {journal} {\bibinfo  {journal} {Physical Review
  Letters}\ }\textbf {\bibinfo {volume} {42}},\ \bibinfo {pages} {1165}
  (\bibinfo {year} {1979})}\BibitemShut {NoStop}%
\bibitem [{\citenamefont {Hebard}\ and\ \citenamefont
  {Fiory}(1980)}]{hebard1980evidence}%
  \BibitemOpen
  \bibfield  {author} {\bibinfo {author} {\bibfnamefont {A.}~\bibnamefont
  {Hebard}}\ and\ \bibinfo {author} {\bibfnamefont {A.}~\bibnamefont {Fiory}},\
  }\bibfield  {title} {\enquote {\bibinfo {title} {Evidence for the
  kosterlitz-thouless transition in thin superconducting aluminum films},}\
  }\href@noop {} {\bibfield  {journal} {\bibinfo  {journal} {Physical Review
  Letters}\ }\textbf {\bibinfo {volume} {44}},\ \bibinfo {pages} {291}
  (\bibinfo {year} {1980})}\BibitemShut {NoStop}%
\bibitem [{\citenamefont {Pargellis}, \citenamefont {Green},\ and\
  \citenamefont {Yurke}(1994)}]{pargellis1994planar}%
  \BibitemOpen
  \bibfield  {author} {\bibinfo {author} {\bibfnamefont {A.~N.}\ \bibnamefont
  {Pargellis}}, \bibinfo {author} {\bibfnamefont {S.}~\bibnamefont {Green}}, \
  and\ \bibinfo {author} {\bibfnamefont {B.}~\bibnamefont {Yurke}},\ }\bibfield
   {title} {\enquote {\bibinfo {title} {Planar xy-model dynamics in a nematic
  liquid crystal system},}\ }\href@noop {} {\bibfield  {journal} {\bibinfo
  {journal} {Physical Review E}\ }\textbf {\bibinfo {volume} {49}},\ \bibinfo
  {pages} {4250} (\bibinfo {year} {1994})}\BibitemShut {NoStop}%
\bibitem [{\citenamefont {Halperin}\ and\ \citenamefont
  {Nelson}(1978)}]{halperin1978theory}%
  \BibitemOpen
  \bibfield  {author} {\bibinfo {author} {\bibfnamefont {B.}~\bibnamefont
  {Halperin}}\ and\ \bibinfo {author} {\bibfnamefont {D.~R.}\ \bibnamefont
  {Nelson}},\ }\bibfield  {title} {\enquote {\bibinfo {title} {Theory of
  two-dimensional melting},}\ }\href@noop {} {\bibfield  {journal} {\bibinfo
  {journal} {Physical Review Letters}\ }\textbf {\bibinfo {volume} {41}},\
  \bibinfo {pages} {121} (\bibinfo {year} {1978})}\BibitemShut {NoStop}%
\bibitem [{\citenamefont {Nelson}\ and\ \citenamefont
  {Halperin}(1979)}]{nelson1979dislocation}%
  \BibitemOpen
  \bibfield  {author} {\bibinfo {author} {\bibfnamefont {D.~R.}\ \bibnamefont
  {Nelson}}\ and\ \bibinfo {author} {\bibfnamefont {B.}~\bibnamefont
  {Halperin}},\ }\bibfield  {title} {\enquote {\bibinfo {title}
  {Dislocation-mediated melting in two dimensions},}\ }\href@noop {} {\bibfield
   {journal} {\bibinfo  {journal} {Physical Review B}\ }\textbf {\bibinfo
  {volume} {19}},\ \bibinfo {pages} {2457} (\bibinfo {year}
  {1979})}\BibitemShut {NoStop}%
\bibitem [{\citenamefont {Nixon}\ \emph {et~al.}(2013)\citenamefont {Nixon},
  \citenamefont {Ronen}, \citenamefont {Friesem},\ and\ \citenamefont
  {Davidson}}]{nixon2013observing}%
  \BibitemOpen
  \bibfield  {author} {\bibinfo {author} {\bibfnamefont {M.}~\bibnamefont
  {Nixon}}, \bibinfo {author} {\bibfnamefont {E.}~\bibnamefont {Ronen}},
  \bibinfo {author} {\bibfnamefont {A.~A.}\ \bibnamefont {Friesem}}, \ and\
  \bibinfo {author} {\bibfnamefont {N.}~\bibnamefont {Davidson}},\ }\bibfield
  {title} {\enquote {\bibinfo {title} {Observing geometric frustration with
  thousands of coupled lasers},}\ }\href {\doibase
  10.1103/PhysRevLett.110.184102} {\bibfield  {journal} {\bibinfo  {journal}
  {Phys. Rev. Lett.}\ }\textbf {\bibinfo {volume} {110}},\ \bibinfo {pages}
  {184102} (\bibinfo {year} {2013})}\BibitemShut {NoStop}%
\bibitem [{\citenamefont {Berloff}\ \emph {et~al.}(2017)\citenamefont
  {Berloff}, \citenamefont {Silva}, \citenamefont {Kalinin}, \citenamefont
  {Askitopoulos}, \citenamefont {Töpfer}, \citenamefont {Cilibrizzi},
  \citenamefont {Langbein},\ and\ \citenamefont
  {Lagoudakis}}]{berloff2017realizing}%
  \BibitemOpen
  \bibfield  {author} {\bibinfo {author} {\bibfnamefont {N.~G.}\ \bibnamefont
  {Berloff}}, \bibinfo {author} {\bibfnamefont {M.}~\bibnamefont {Silva}},
  \bibinfo {author} {\bibfnamefont {K.}~\bibnamefont {Kalinin}}, \bibinfo
  {author} {\bibfnamefont {A.}~\bibnamefont {Askitopoulos}}, \bibinfo {author}
  {\bibfnamefont {J.~D.}\ \bibnamefont {Töpfer}}, \bibinfo {author}
  {\bibfnamefont {P.}~\bibnamefont {Cilibrizzi}}, \bibinfo {author}
  {\bibfnamefont {W.}~\bibnamefont {Langbein}}, \ and\ \bibinfo {author}
  {\bibfnamefont {P.~G.}\ \bibnamefont {Lagoudakis}},\ }\bibfield  {title}
  {\enquote {\bibinfo {title} {Realizing the classical {XY} {Hamiltonian} in
  polariton simulators},}\ }\href {\doibase 10.1038/nmat4971} {\bibfield
  {journal} {\bibinfo  {journal} {Nature Materials}\ }\textbf {\bibinfo
  {volume} {16}},\ \bibinfo {pages} {1120--1126} (\bibinfo {year}
  {2017})}\BibitemShut {NoStop}%
\bibitem [{\citenamefont {Parto}\ \emph {et~al.}(2020)\citenamefont {Parto},
  \citenamefont {Hayenga}, \citenamefont {Marandi}, \citenamefont
  {Christodoulides},\ and\ \citenamefont {Khajavikhan}}]{parto2020realizing}%
  \BibitemOpen
  \bibfield  {author} {\bibinfo {author} {\bibfnamefont {M.}~\bibnamefont
  {Parto}}, \bibinfo {author} {\bibfnamefont {W.}~\bibnamefont {Hayenga}},
  \bibinfo {author} {\bibfnamefont {A.}~\bibnamefont {Marandi}}, \bibinfo
  {author} {\bibfnamefont {D.~N.}\ \bibnamefont {Christodoulides}}, \ and\
  \bibinfo {author} {\bibfnamefont {M.}~\bibnamefont {Khajavikhan}},\
  }\bibfield  {title} {\enquote {\bibinfo {title} {Realizing spin
  {Hamiltonians} in nanoscale active photonic lattices},}\ }\href {\doibase
  10.1038/s41563-020-0635-6} {\bibfield  {journal} {\bibinfo  {journal} {Nature
  Materials}\ } (\bibinfo {year} {2020}),\
  10.1038/s41563-020-0635-6}\BibitemShut {NoStop}%
\bibitem [{\citenamefont {Honari-Latifpour}\ and\ \citenamefont
  {Miri}(2020)}]{honari_2020}%
  \BibitemOpen
  \bibfield  {author} {\bibinfo {author} {\bibfnamefont {M.}~\bibnamefont
  {Honari-Latifpour}}\ and\ \bibinfo {author} {\bibfnamefont {M.-A.}\
  \bibnamefont {Miri}},\ }\bibfield  {title} {\enquote {\bibinfo {title}
  {Mapping the $xy$ hamiltonian onto a network of coupled lasers},}\ }\href
  {\doibase 10.1103/PhysRevResearch.2.043335} {\bibfield  {journal} {\bibinfo
  {journal} {Phys. Rev. Research}\ }\textbf {\bibinfo {volume} {2}},\ \bibinfo
  {pages} {043335} (\bibinfo {year} {2020})}\BibitemShut {NoStop}%
\bibitem [{\citenamefont {Ding}\ \emph {et~al.}(2019)\citenamefont {Ding},
  \citenamefont {Belykh}, \citenamefont {Marandi},\ and\ \citenamefont
  {Miri}}]{ding2019dispersive}%
  \BibitemOpen
  \bibfield  {author} {\bibinfo {author} {\bibfnamefont {J.}~\bibnamefont
  {Ding}}, \bibinfo {author} {\bibfnamefont {I.}~\bibnamefont {Belykh}},
  \bibinfo {author} {\bibfnamefont {A.}~\bibnamefont {Marandi}}, \ and\
  \bibinfo {author} {\bibfnamefont {M.-A.}\ \bibnamefont {Miri}},\ }\bibfield
  {title} {\enquote {\bibinfo {title} {Dispersive versus dissipative coupling
  for frequency synchronization in lasers},}\ }\href {\doibase
  10.1103/PhysRevApplied.12.054039} {\bibfield  {journal} {\bibinfo  {journal}
  {Phys. Rev. Applied}\ }\textbf {\bibinfo {volume} {12}},\ \bibinfo {pages}
  {054039} (\bibinfo {year} {2019})}\BibitemShut {NoStop}%
\bibitem [{\citenamefont {Ding}\ and\ \citenamefont
  {Miri}(2019)}]{ding2019mode}%
  \BibitemOpen
  \bibfield  {author} {\bibinfo {author} {\bibfnamefont {J.}~\bibnamefont
  {Ding}}\ and\ \bibinfo {author} {\bibfnamefont {M.-A.}\ \bibnamefont
  {Miri}},\ }\bibfield  {title} {\enquote {\bibinfo {title} {Mode
  discrimination in dissipatively coupled laser arrays},}\ }\href@noop {}
  {\bibfield  {journal} {\bibinfo  {journal} {Optics letters}\ }\textbf
  {\bibinfo {volume} {44}},\ \bibinfo {pages} {5021--5024} (\bibinfo {year}
  {2019})}\BibitemShut {NoStop}%
\bibitem [{\citenamefont {Haus}(1984)}]{haus1984waves}%
  \BibitemOpen
  \bibfield  {author} {\bibinfo {author} {\bibfnamefont {H.}~\bibnamefont
  {Haus}},\ }\bibfield  {title} {\enquote {\bibinfo {title} {Waves and fields
  in optoelectronics.}}\ }\href@noop {} {\bibfield  {journal} {\bibinfo
  {journal} {PRENTICE-HALL, INC., ENGLEWOOD CLIFFS, NJ 07632, USA, 1984, 402}\
  } (\bibinfo {year} {1984})}\BibitemShut {NoStop}%
\bibitem [{\citenamefont {Suh}, \citenamefont {Wang},\ and\ \citenamefont
  {Fan}(2004)}]{suh2004temporal}%
  \BibitemOpen
  \bibfield  {author} {\bibinfo {author} {\bibfnamefont {W.}~\bibnamefont
  {Suh}}, \bibinfo {author} {\bibfnamefont {Z.}~\bibnamefont {Wang}}, \ and\
  \bibinfo {author} {\bibfnamefont {S.}~\bibnamefont {Fan}},\ }\bibfield
  {title} {\enquote {\bibinfo {title} {Temporal coupled-mode theory and the
  presence of non-orthogonal modes in lossless multimode cavities},}\
  }\href@noop {} {\bibfield  {journal} {\bibinfo  {journal} {IEEE Journal of
  Quantum Electronics}\ }\textbf {\bibinfo {volume} {40}},\ \bibinfo {pages}
  {1511--1518} (\bibinfo {year} {2004})}\BibitemShut {NoStop}%
\bibitem [{\citenamefont {Tredicce}\ \emph {et~al.}(1985)\citenamefont
  {Tredicce}, \citenamefont {Arecchi}, \citenamefont {Lippi},\ and\
  \citenamefont {Puccioni}}]{tredicce1985instabilities}%
  \BibitemOpen
  \bibfield  {author} {\bibinfo {author} {\bibfnamefont {J.~R.}\ \bibnamefont
  {Tredicce}}, \bibinfo {author} {\bibfnamefont {F.~T.}\ \bibnamefont
  {Arecchi}}, \bibinfo {author} {\bibfnamefont {G.~L.}\ \bibnamefont {Lippi}},
  \ and\ \bibinfo {author} {\bibfnamefont {G.~P.}\ \bibnamefont {Puccioni}},\
  }\bibfield  {title} {\enquote {\bibinfo {title} {Instabilities in lasers with
  an injected signal},}\ }\href@noop {} {\bibfield  {journal} {\bibinfo
  {journal} {JOSA B}\ }\textbf {\bibinfo {volume} {2}},\ \bibinfo {pages}
  {173--183} (\bibinfo {year} {1985})}\BibitemShut {NoStop}%
\bibitem [{\citenamefont {Fabiny}\ \emph {et~al.}(1993)\citenamefont {Fabiny},
  \citenamefont {Colet}, \citenamefont {Roy},\ and\ \citenamefont
  {Lenstra}}]{fabiny1993coherence}%
  \BibitemOpen
  \bibfield  {author} {\bibinfo {author} {\bibfnamefont {L.}~\bibnamefont
  {Fabiny}}, \bibinfo {author} {\bibfnamefont {P.}~\bibnamefont {Colet}},
  \bibinfo {author} {\bibfnamefont {R.}~\bibnamefont {Roy}}, \ and\ \bibinfo
  {author} {\bibfnamefont {D.}~\bibnamefont {Lenstra}},\ }\bibfield  {title}
  {\enquote {\bibinfo {title} {Coherence and phase dynamics of spatially
  coupled solid-state lasers},}\ }\href@noop {} {\bibfield  {journal} {\bibinfo
   {journal} {Physical Review A}\ }\textbf {\bibinfo {volume} {47}},\ \bibinfo
  {pages} {4287} (\bibinfo {year} {1993})}\BibitemShut {NoStop}%
\bibitem [{\citenamefont {Ohtsubo}(2012)}]{ohtsubo2012semiconductor}%
  \BibitemOpen
  \bibfield  {author} {\bibinfo {author} {\bibfnamefont {J.}~\bibnamefont
  {Ohtsubo}},\ }\href@noop {} {\emph {\bibinfo {title} {Semiconductor lasers:
  stability, instability and chaos}}},\ Vol.\ \bibinfo {volume} {111}\
  (\bibinfo  {publisher} {Springer},\ \bibinfo {year} {2012})\BibitemShut
  {NoStop}%
\bibitem [{\citenamefont {Mandel}\ and\ \citenamefont
  {Wolf}(1995)}]{mandel1995optical}%
  \BibitemOpen
  \bibfield  {author} {\bibinfo {author} {\bibfnamefont {L.}~\bibnamefont
  {Mandel}}\ and\ \bibinfo {author} {\bibfnamefont {E.}~\bibnamefont {Wolf}},\
  }\href@noop {} {\emph {\bibinfo {title} {Optical coherence and quantum
  optics}}}\ (\bibinfo  {publisher} {Cambridge University Press},\ \bibinfo
  {year} {1995})\BibitemShut {NoStop}%
\bibitem [{\citenamefont {Mahler}\ \emph {et~al.}(2019)\citenamefont {Mahler},
  \citenamefont {Pal}, \citenamefont {Tradonsky}, \citenamefont {Chriki},
  \citenamefont {Friesem},\ and\ \citenamefont
  {Davidson}}]{mahler2019dynamics}%
  \BibitemOpen
  \bibfield  {author} {\bibinfo {author} {\bibfnamefont {S.}~\bibnamefont
  {Mahler}}, \bibinfo {author} {\bibfnamefont {V.}~\bibnamefont {Pal}},
  \bibinfo {author} {\bibfnamefont {C.}~\bibnamefont {Tradonsky}}, \bibinfo
  {author} {\bibfnamefont {R.}~\bibnamefont {Chriki}}, \bibinfo {author}
  {\bibfnamefont {A.~A.}\ \bibnamefont {Friesem}}, \ and\ \bibinfo {author}
  {\bibfnamefont {N.}~\bibnamefont {Davidson}},\ }\bibfield  {title} {\enquote
  {\bibinfo {title} {Dynamics of dissipative topological defects in coupled
  phase oscillators},}\ }\href@noop {} {\bibfield  {journal} {\bibinfo
  {journal} {Journal of Physics B: Atomic, Molecular and Optical Physics}\
  }\textbf {\bibinfo {volume} {52}},\ \bibinfo {pages} {205401} (\bibinfo
  {year} {2019})}\BibitemShut {NoStop}%
\bibitem [{\citenamefont {Aranson}\ and\ \citenamefont
  {Kramer}(2002)}]{aranson2002world}%
  \BibitemOpen
  \bibfield  {author} {\bibinfo {author} {\bibfnamefont {I.~S.}\ \bibnamefont
  {Aranson}}\ and\ \bibinfo {author} {\bibfnamefont {L.}~\bibnamefont
  {Kramer}},\ }\bibfield  {title} {\enquote {\bibinfo {title} {The world of the
  complex ginzburg-landau equation},}\ }\href@noop {} {\bibfield  {journal}
  {\bibinfo  {journal} {Reviews of modern physics}\ }\textbf {\bibinfo {volume}
  {74}},\ \bibinfo {pages} {99} (\bibinfo {year} {2002})}\BibitemShut {NoStop}%
\bibitem [{\citenamefont {Hagan}(1982)}]{hagan1982spiral}%
  \BibitemOpen
  \bibfield  {author} {\bibinfo {author} {\bibfnamefont {P.~S.}\ \bibnamefont
  {Hagan}},\ }\bibfield  {title} {\enquote {\bibinfo {title} {Spiral waves in
  reaction-diffusion equations},}\ }\href@noop {} {\bibfield  {journal}
  {\bibinfo  {journal} {SIAM journal on applied mathematics}\ }\textbf
  {\bibinfo {volume} {42}},\ \bibinfo {pages} {762--786} (\bibinfo {year}
  {1982})}\BibitemShut {NoStop}%
\bibitem [{\citenamefont {Mermin}\ and\ \citenamefont
  {Wagner}(1966)}]{mermin1966absence}%
  \BibitemOpen
  \bibfield  {author} {\bibinfo {author} {\bibfnamefont {N.~D.}\ \bibnamefont
  {Mermin}}\ and\ \bibinfo {author} {\bibfnamefont {H.}~\bibnamefont
  {Wagner}},\ }\bibfield  {title} {\enquote {\bibinfo {title} {Absence of
  ferromagnetism or antiferromagnetism in one-or two-dimensional isotropic
  heisenberg models},}\ }\href@noop {} {\bibfield  {journal} {\bibinfo
  {journal} {Physical Review Letters}\ }\textbf {\bibinfo {volume} {17}},\
  \bibinfo {pages} {1133} (\bibinfo {year} {1966})}\BibitemShut {NoStop}%
\bibitem [{\citenamefont {Kosterlitz}\ and\ \citenamefont
  {Thouless}(1973)}]{kosterlitz1973j}%
  \BibitemOpen
  \bibfield  {author} {\bibinfo {author} {\bibfnamefont {J.}~\bibnamefont
  {Kosterlitz}}\ and\ \bibinfo {author} {\bibfnamefont {D.}~\bibnamefont
  {Thouless}},\ }\bibfield  {title} {\enquote {\bibinfo {title} {J. phys. c:
  Solid state phys.}}\ }\href@noop {} {\  (\bibinfo {year} {1973})}\BibitemShut
  {NoStop}%
\bibitem [{\citenamefont {Kosterlitz}(1974)}]{kosterlitz1974critical}%
  \BibitemOpen
  \bibfield  {author} {\bibinfo {author} {\bibfnamefont {J.}~\bibnamefont
  {Kosterlitz}},\ }\bibfield  {title} {\enquote {\bibinfo {title} {The critical
  properties of the two-dimensional xy model},}\ }\href@noop {} {\bibfield
  {journal} {\bibinfo  {journal} {Journal of Physics C: Solid State Physics}\
  }\textbf {\bibinfo {volume} {7}},\ \bibinfo {pages} {1046} (\bibinfo {year}
  {1974})}\BibitemShut {NoStop}%
\bibitem [{\citenamefont {Lee}\ \emph {et~al.}(2010)\citenamefont {Lee},
  \citenamefont {Tam}, \citenamefont {Refael}, \citenamefont {Rogers},\ and\
  \citenamefont {Cross}}]{lee2010vortices}%
  \BibitemOpen
  \bibfield  {author} {\bibinfo {author} {\bibfnamefont {T.~E.}\ \bibnamefont
  {Lee}}, \bibinfo {author} {\bibfnamefont {H.}~\bibnamefont {Tam}}, \bibinfo
  {author} {\bibfnamefont {G.}~\bibnamefont {Refael}}, \bibinfo {author}
  {\bibfnamefont {J.~L.}\ \bibnamefont {Rogers}}, \ and\ \bibinfo {author}
  {\bibfnamefont {M.}~\bibnamefont {Cross}},\ }\bibfield  {title} {\enquote
  {\bibinfo {title} {Vortices and the entrainment transition in the
  two-dimensional kuramoto model},}\ }\href@noop {} {\bibfield  {journal}
  {\bibinfo  {journal} {Physical Review E}\ }\textbf {\bibinfo {volume} {82}},\
  \bibinfo {pages} {036202} (\bibinfo {year} {2010})}\BibitemShut {NoStop}%
\end{thebibliography}%

\end{document}